\documentclass[letterpaper]{article} 
\usepackage{aaai2026}  
\usepackage{times}  
\usepackage{helvet}  
\usepackage{courier}  
\usepackage[hyphens]{url}  
\usepackage{graphicx} 
\usepackage{natbib}  
\usepackage{caption} 
\usepackage{algorithm}
\usepackage{booktabs}
\usepackage{caption}
\usepackage{array}
\usepackage{adjustbox}
\usepackage{multirow}
\usepackage{amsmath}
\usepackage{amssymb}
\usepackage{amsfonts}
\usepackage{algorithmicx}
\usepackage{algpseudocode}
\usepackage[table]{xcolor}
\usepackage{subcaption}
\usepackage{graphicx}
\usepackage{newfloat}
\usepackage{listings}
\DeclareCaptionStyle{ruled}{labelfont=normalfont,labelsep=colon,strut=off} 
\floatstyle{ruled}
\newfloat{listing}{tb}{lst}{}
\floatname{listing}{Listing}
\title{Neural-Augmented Kelvinlet for Real-Time Soft Tissue Deformation Modeling}

\author {
    Ashkan Shahbazi\textsuperscript{\rm 1}\thanks{Equal contribution},
    Kyvia Pereira\textsuperscript{\rm 1}\footnotemark[1],
    Jon S. Heiselman\textsuperscript{\rm 1},
    Elaheh Akbari\textsuperscript{\rm 1},
    Annie C. Benson\textsuperscript{\rm 1},\\
    Sepehr Seifi\textsuperscript{\rm 1},
    Xinyuan Liu\textsuperscript{\rm 1},
    Garrison Lawrence Horswill Johnston\textsuperscript{\rm 1},
    Jie Ying Wu\textsuperscript{\rm 1},\\
    Nabil Simaan\textsuperscript{\rm 1},
    Michael Miga\textsuperscript{\rm 1},
    Soheil Kolouri\textsuperscript{\rm 1}
}
\affiliations{
{\fontsize{10}{12}\selectfont \normalfont
\textsuperscript{\rm 1}Vanderbilt University, USA\\
\{ashkan.shahbazi, kyvia.pereira, jon.s.heiselman, elaheh.akbari, annie.c.benson,\\
sepehr.seifi, xinyuan.liu, garrison.l.johnston, jieying.wu,
nabil.simaan, michael.i.miga, soheil.kolouri\}@vanderbilt.edu
}}

\begin{document}

\maketitle

\begin{abstract}
Accurate and efficient modeling of soft-tissue interactions is fundamental for advancing surgical simulation, surgical robotics, and model-based surgical automation. To achieve real-time latency, classical Finite Element Method (FEM) solvers are often replaced with neural approximations; however, naively training such models in a fully data-driven manner without incorporating physical priors frequently leads to poor generalization and physically implausible predictions. We present a novel physics-informed neural simulation framework that enables real-time prediction of soft-tissue deformations under complex single- and multi-grasper interactions. Our approach integrates Kelvinlet-based analytical priors with large-scale FEM data, capturing both linear and nonlinear tissue responses. This hybrid design improves predictive accuracy and physical plausibility across diverse neural architectures while maintaining the low-latency performance required for interactive applications. We validate our method on challenging surgical manipulation tasks involving standard laparoscopic grasping tools, demonstrating substantial improvements in deformation fidelity and temporal stability over existing baselines. These results establish Kelvinlet-augmented learning as a principled and computationally efficient paradigm for real-time, physics-aware soft-tissue simulation in surgical AI.
\end{abstract}

\begin{links}
    \link{Code and data}{https://github.com/mint-vu/neural-augmented-kelvinlet-leap}
\end{links}

\section{Introduction}


Soft tissue deformation modeling is essential in computational biomechanics for surgical simulation, image registration, and robotic-assisted interventions. Accurate tissue simulation enables surgical planning, image-guided predictions of soft-tissue changes, and precise haptic feedback for robotics. Traditional approaches rely on physics-based methods grounded in continuum mechanics, such as the Finite Element Method (FEM), Boundary Element Method (BEM), and Tensor Mass Model (TMM). While highly accurate and physically realistic, these methods are often computationally prohibitive for real-time applications \cite{ALLARD2012281,zhang2019soft}. To reduce computational cost, approximate methods such as mass-spring systems \cite{nedel1998deformable}, mesh-free methods \cite{belytschko1996meshfree}, and particle-based methods \cite{liu2015deformation} are frequently used. However, these approximations often sacrifice realism, creating a significant simulation-to-reality gap that limits their practicality in high-fidelity applications.

Recent advances in deep learning and neural operators have enabled fully data-driven methods that approximate soft tissue deformations with a single forward pass of a learned model. Convolutional neural networks (CNNs), graph neural networks (GNNs), and Transformers \cite{wu2020leveraging,wu2021learning,pfaff2021learning} have demonstrated the ability to capture complex deformation patterns from simulated or real-world data, enabling real-time tissue simulation. Despite these advances, fully data-driven approaches face two fundamental limitations: (a) limited generalization to unseen force distributions and boundary conditions, which results in unreliable extrapolation, and (b) a tendency to produce physically inconsistent deformations, particularly when trained on datasets of limited size or diversity. These shortcomings highlight the need for hybrid approaches that combine the expressiveness of neural models with the robustness of physics-based priors.

To address these challenges, considerable efforts have focused on integrating physics priors into data-driven machine learning models \cite{karniadakis2021physics}. Broadly, these methods make learning algorithms physics-informed by introducing: (a) observational biases, such as data augmentations that embed physical constraints; (b) inductive biases in the model architecture, for example, using group-invariant or equivariant neural networks \cite{bronstein2017geometric}; and (c) learning biases through the design of loss functions that explicitly enforce physical laws. Despite this progress, achieving the necessary low latency for real-time simulation while maintaining a high-performing physics-informed model with appropriately chosen biases remains a challenging and largely open problem.

\textbf{In this paper}, we model soft tissue using a volumetric mesh and address the problem of predicting the displacement of the entire mesh given the observed displacements at a limited set of interaction points on its surface. This is analogous to determining how the entire soft tissue will deform when graspers engage multiple surface locations and are moved to specified target positions.
While several prior studies have explored similar setup and demonstrated that neural networks can approximate the solutions produced by high-fidelity physics-based solvers such as FEM \cite{pfeiffer2019learning}, these approaches often lack strong physical priors and can suffer from reduced generalization and physical inconsistency. To overcome these limitations, we propose a physics-informed neural framework that integrates analytical elastic priors to enhance both accuracy and physical plausibility.

We introduce a physics-informed neural framework that leverages Kelvinlets \cite{goes2017kelvinlet}, i.e., the closed-form Green’s function solutions to the Navier–Cauchy equations, as an analytically tractable and computationally efficient prior. Kelvinlets provide a fast, low-order approximation of elastic deformation that captures the fundamental physics of linear elasticity. We build on this prior through two complementary hybrid strategies: (a) \emph{residual learning:} using Kelvinlets to generate a baseline displacement field and training a neural network to model the residual error; and (b) \emph{physics regularization:} incorporating Kelvinlets as a soft constraint within the learning objective of a neural network that predicts the full deformation field.
By explicitly coupling the expressiveness of neural networks with the efficiency and physical grounding of Kelvinlets, our approach achieves improved accuracy, generalization, and physical plausibility for soft tissue deformation modeling.

To establish a unified testbed for evaluating our approach alongside prior methods, we generated a finite element dataset of liver deformations using both a linear model without preload and a nonlinear model with gravity-induced preload. We show that our proposed \emph{Neural-Augmented Kelvinlet} framework surpasses purely data-driven models in generalization, physical realism, and learning efficiency, ultimately enabling real-time, high-fidelity modeling of soft tissue deformations involving multiple graspers.\\


\noindent\textbf{Contributions.} Our key contributions are:
\begin{enumerate}    
    \item We present a dataset of 20,800 FEM solutions for data-driven soft tissue modeling, with 10,400 linear and 10,400 nonlinear simulations. It includes 5,000 single-grasper and 5,400 multi-grasper cases.
    \item We provide a GPU-accelerated Kelvinlet PyTorch implementation that enables efficient training and real-time inference, making physics-informed deformation modeling accessible for interactive applications.
    \item We introduce two hybrid approaches that integrate Kelvinlets into data-driven soft tissue deformation modeling and conduct extensive experiments to demonstrate their superiority over purely physics-informed and data-driven baselines across diverse neural architectures.
\end{enumerate}

\section{Related Work}


Classical approaches based on the Finite Element Method (FEM) provide high-fidelity simulations of soft tissue deformation but are often computationally prohibitive for real-time surgical applications. To address this limitation, model reduction techniques such as Proper Orthogonal Decomposition (POD) have been proposed to accelerate FEM without sacrificing physical realism. For example, Lau et al. \cite{liu2025data} employ a reduced-order linear FEM with modal warping to achieve sub-15 ms inference for liver deformation, enabling intraoperative use. Complementarily, Hu and Desai \cite{10.1007/978-3-540-25968-8_4} characterize nonlinear liver material behavior through indentation experiments and validate Local Effective Moduli (LEM) estimates using hyperelastic FEM models, laying foundational work for realistic soft-tissue mechanics under large strains.

More recently, physics-informed neural networks (PINNs) have emerged as a data-efficient alternative that integrates governing equations such as elasticity or dynamics directly into the learning objective. Hu et al. \cite{hu2025realtime} introduce a PINN that encodes static linear elasticity to predict 3D stress and strain fields of liver tissue under surgical tool interaction, achieving millisecond-scale inference while retaining physical plausibility. Extending this paradigm, Nguyen-Le et al. \cite{NGUYENLE2025111217} propose a PINODE framework that augments neural dynamics with mass-spring priors, enabling extrapolation from sparse motion data. These methods improve interpretability and generalization but often require complex training and remain constrained by solver latency or limited scalability.

In contrast, purely data-driven surrogate models have been proposed to directly regress deformation fields from control variables. Ke et al. \cite{camara2016soft} train a feedforward network to predict liver surface displacements conditioned on retractor pose, achieving sub-millisecond inference. While fast and simple, such models lack physical constraints, which can reduce reliability and robustness in unseen conditions.

On the learning-based front, mesh-based models such as PhysGNN leverage FEM supervision to train graph neural networks for real-time deformation prediction in neurosurgery \cite{salehi2022physgnn}. PEGNN \cite{saleh2024physics} extends this with contact-aware hybrid graph reasoning, enabling accurate predictions across surgical and manipulation scenarios. Meanwhile, SeeSaw \cite{10598377} employs stereo video and self-supervised learning to estimate sparse 3D deformations using a GNN encoder. Zhao et al. \cite{10630572} introduce a diffeomorphic mapping framework that enforces cycle consistency, capturing invertible tissue motion. These approaches show strong empirical performance but may suffer from limited interpretability or reliance on dense supervision.

Neural architectures such as U-Net \cite{mendizabal2020simulation}, Set Attention Block (SAB) \cite{lee2019set}, and GraphGPS \cite{rampasek2022GPS} offer modular and scalable backbones for modeling deformation fields across grid-based, set-based, and graph-structured data. While effective in capturing spatial patterns and long-range dependencies, these models are generally agnostic to the underlying physics of soft tissue behavior and typically lack inductive biases grounded in biomechanics, such as force equilibrium, boundary conditions, or constitutive material priors.

Our work builds upon these foundations by integrating physically motivated priors, specifically Kelvinlet-based elastic solutions, into neural architectures through both residual learning and regularization. By embedding closed-form Green’s function responses from linear elasticity into the training process, our method encourages physically consistent deformation predictions even in data-sparse regimes. This hybrid approach improves generalization, improves physical plausibility, and remains compatible with a range of backbone architectures, while preserving the low latency performance required for real-time surgical simulation and interactive applications.

\section{Large-Scale FEM Data}


\begin{figure}[t!]
    \centering
    \includegraphics[width=\columnwidth]{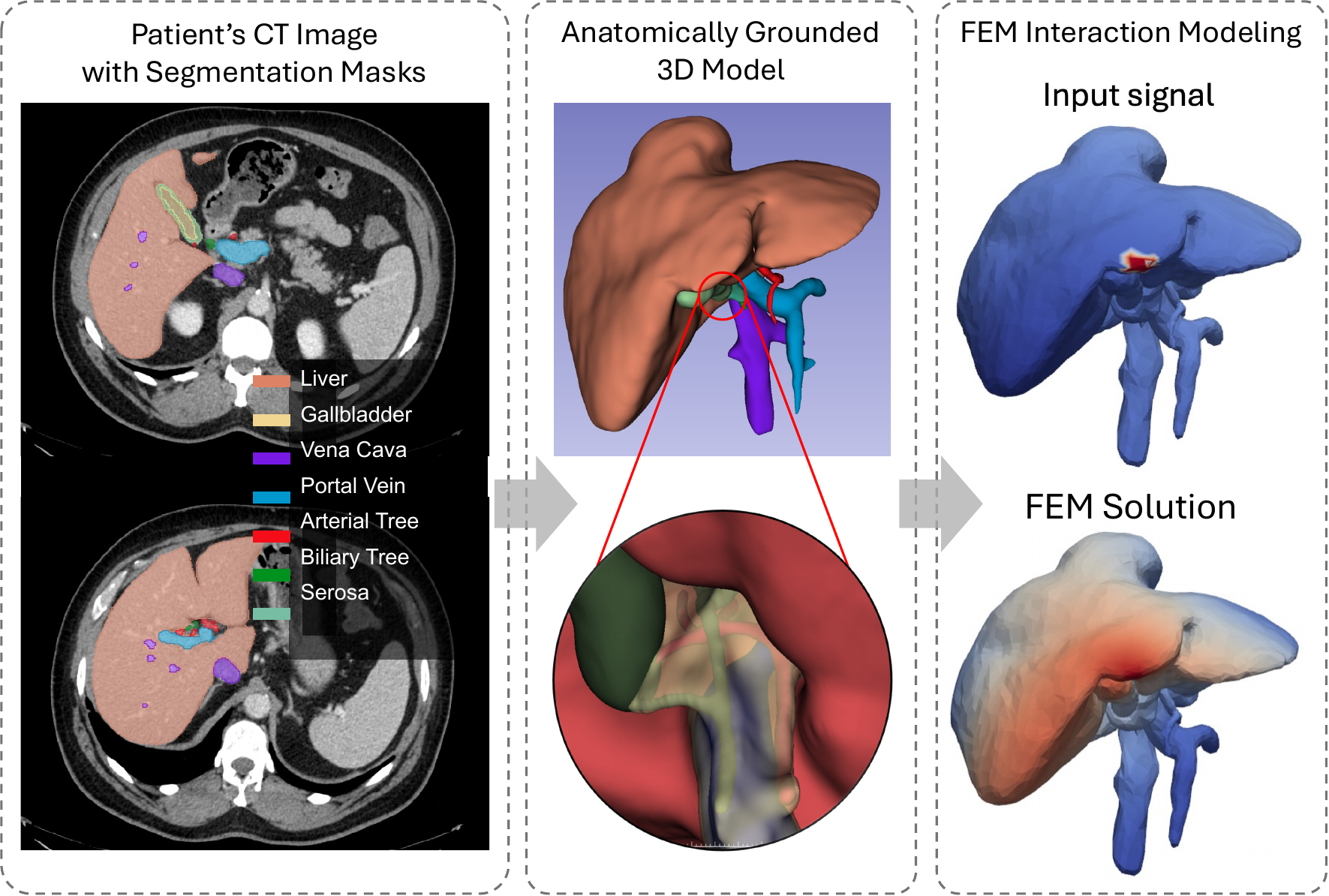}
    \caption{Overview of the large-scale FEM data generation pipeline. A patient’s abdominal CT scan is first anatomically segmented (left), followed by the generation of a 3D linear tetrahedral mesh (middle). Diverse interactions are then applied to the mesh, and FEM simulations are performed to produce training data for ML models (right). The right panel shows the norms of the input (interactions) and the resulting output displacements (predicted global effects).}
    \label{fig:data}
    \vspace{-0.25in}
\end{figure}

Neural network–based solutions require training on large-scale finite element (FE) simulation datasets. In this work, we focus on laparoscopic cholecystectomy and construct an FE model from an abdominal CT image and its segmentation masks, converting them into a 3D linear tetrahedral mesh with 10,400 nodes and 44,045 tetrahedra using \texttt{gmsh} \cite{gmsh2009reference}. As part of our contributions, we release this large-scale dataset; the details of its creation are provided below.

The bare area of the liver, where it contacts the inferior vena cava and diaphragm, and the distal ends of the ducts and veins are fixed (non-deformable). Tool interaction regions are selected to simulate the initial surgical step of lifting the liver lobes to expose the gallbladder and cystic ducts. These regions are identified by clustering nodes along the liver ridge, based on surgical video analysis and input from our surgical collaborator. FE simulations are performed using the open-source getFEM package \cite{getFEM2020}, applying tool-imposed displacements at those predefined clustered regions. This pipeline is illustrated in Figure \ref{fig:data}.

In our FE model, nodes with restricted motion or tool-imposed displacements are flagged. These boundary conditions are encoded as binary features in the neural network (1 for constrained or controlled, 0 otherwise), allowing the model to distinguish between free and constrained regions. 



Assuming a quasi-static environment, our solutions satisfy equilibrium by ensuring that the total potential energy is not only stationary but is also a minimum for permissible displacements, and is given by: $\int_{\Omega} \frac{\boldsymbol{W}(\boldsymbol{E})}{\partial \boldsymbol{\boldsymbol{E}}} : \delta \boldsymbol{E}\, dV + \int_{\Omega} \rho \boldsymbol{g} \cdot \delta \mathbf{u}\, dV = 0$, where $\boldsymbol{W}(\boldsymbol{E})$ is the strain energy density, $\Omega$ is the volume of the model, $\delta \mathbf{u}$ is the variation of the displacement field, $\delta \boldsymbol{E}$ is the variation of the Green-Lagrange strain tensor, $\rho$ is the density of the material (1000 $kg / m^3$), and $\boldsymbol{g}$ is the acceleration due to gravity that points downward along the y-axis (9.81 $m /s^2$). For large displacements, the Green-Lagrange strain tensor $\boldsymbol{E}$ depends on the unknown displacement field $\mathbf{u}$ and is defined as:
\begin{equation}
\label{eq:deform_tensor}
\boldsymbol{E} = \frac{1}{2} \left(\nabla \mathbf{u} + \nabla \mathbf{u}^T + \nabla \mathbf{u}^T \nabla \mathbf{u} \right).
\end{equation}
This equation relates to the Cauchy-Green deformation tensor $\boldsymbol{C}$ by: $\boldsymbol{C} = 2\boldsymbol{E} - \mathbf{I}$, where $\mathbf{I}$ is the identity tensor.

For our nonlinear simulations, we use a two-parameter Mooney-Rivlin model \cite{rivlin1948large}, where the strain energy density can be written as functions of the Cauchy-Green deformation tensor invariants ($\boldsymbol{I_i(C)}$, with $i=1,2,3$) \cite{Hackett2018}: 
{
\begin{multline} \label{eq:strain-energy}
\boldsymbol{W} = C_{01} \left( I_1(\mathbf{C}) \cdot I_3(\mathbf{C})^{-1/3} - 3 \right) \\
+ C_{10} \left( I_1(\mathbf{C}) \cdot I_3(\mathbf{C})^{-2/3} - 3 \right),
\end{multline}
}
where $C_{10} = 1.62$ kPa and $C_{01} = 1.97$ kPa are constitutive constants derived from ex-vivo porcine liver experiments \cite{SPIE2025}. 

For our linear simulations, we simulate an elastic response in the absence of gravitational load ($\boldsymbol{g} = 0$). These solutions serve as baseline data for tissue deformation under small elastic input displacements, where the term $\nabla \mathbf{u}^T \nabla \mathbf{u}$ in Eq.~\eqref{eq:deform_tensor} is neglected. Assuming isotropic material properties, the strain energy density $\boldsymbol{W}$ in terms of shear modulus $\mu$ and Poisson's ratio $\nu$  expressed as:
\begin{equation}
\boldsymbol{W}= \left( \frac{\mu \nu}{1 - 2\nu} \right) \mathrm{Tr}(\boldsymbol{E})^2 +\mu \mathrm{Tr}(\boldsymbol{E}^2).
\end{equation}
In these linear simulations, we assign a Poisson's ratio of 0.45 for all organs \cite{chen_youngs_1996}. The shear modulus values are also set as follows: 0.03 kPa for the fascia \cite{akhmanova_physical_2015}, 0.34 kPa for the gallbladder \cite{nisansala_stiffness_2016}, 0.69 kPa for the gallbladder wall and liver \cite{kim_gallbladder_2013}, and 1.07 kPa for the remaining organs \cite{george_influence_2018}.

\subsubsection{Training Distribution, \( p \).}
Given a mesh partitioned into regions \( \Omega_i \) informed by cholecystectomy surgery, nodes \( \mathbf{x}_s \) are sampled from a mixture of Generalized von Mises-Fisher (vMF) distributions (i.e., radial densities on the surface mesh) centered in these regions:
\begin{equation}
    p(\mathbf{x}_s) = \sum_{i} g(\Omega_i) f_{\text{vMF}}(\mathbf{x}_s \mid \boldsymbol{\mu}_i, \kappa_i),
    \label{eq:px}
\end{equation}
where \( g(\Omega_i) \) is the prior-informed probability of selecting \( \Omega_i \), \( \boldsymbol{\mu}_i \) indicates the center of \( \Omega_i \), and \( \kappa_i \) is the spread of the vMF, which depends on the size of \( \Omega_i \). The displacements \( \mathbf{u}_s \in \mathbb{R}^3 \) are independently sampled from a uniform distribution over a cuboid domain \( \mathcal{C} \subset \mathbb{R}^3 \), defined by:
\[
x \in (x_1, x_2),\quad y \in (y_1, y_2),\quad z \in (z_1, z_2),
\]
with total volume \( V = (x_2 - x_1)(y_2 - y_1)(z_2 - z_1) \). This ensures that node sampling is spatially structured according to surgical priors, while displacements remain isotropic within a bounded range.

For single grasping, ten regions \( \Omega_i \) around the liver edge were selected, each with 250 samples. For multiple graspers, six pairwise combinations of five regions \( \Omega_i \) were considered, with 30 samples per region. The displacement bounds are \( x \in (-90, 50) \) mm, \( y \in (-90, 50) \) mm, and \( z \in (-60, 90) \) mm.

\subsubsection{Distribution for Multiple Graspers.}
For multiple interactions (e.g., two graspers), we define the joint distribution as
\(
p(\mathbf{x}_1, \mathbf{x}_2) = p(\mathbf{x}_1)\, p(\mathbf{x}_2 \mid \mathbf{x}_1),
\)
where \( p(\mathbf{x}_1) \) follows Eq.~\eqref{eq:px}. To ensure that \( \mathbf{x}_2 \) is not sampled from the same region \( \Omega_k \) as \( \mathbf{x}_1 \), we modify the region probabilities \( g(\Omega_i) \) by setting \( g(\Omega_k) = 0 \), followed by renormalizing the remaining weights \( g(\Omega_i) \) for \( i \ne k \) based on surgical priors. The conditional distribution \( p(\mathbf{x}_2 \mid \mathbf{x}_1) \) is then defined using these updated region weights.

\subsubsection{Regularization Distribution, \( q \).}
For Kelvinlet regularization, interaction points \( \mathbf{x}_s \) are sampled uniformly from the surface nodes. The corresponding displacements \( \mathbf{u}_s \in \mathbb{R}^3 \) are then sampled uniformly from a cone centered around the surface normal at \( \mathbf{x}_s \), with an opening angle of \( \alpha_{\max} \).

Having detailed the data generation process, we now turn to our proposed neural augmented Kelvinlet solution.

\section{Neural-Augmented Kelvinlet}

In this section, we present the Kelvinlets prior and our residual learning and regularization frameworks.

\subsection{Navier-Cauchy Eqs and Green's Function Solution}

\begin{figure*}[ht!]
    \centering
    \includegraphics[width=\linewidth]{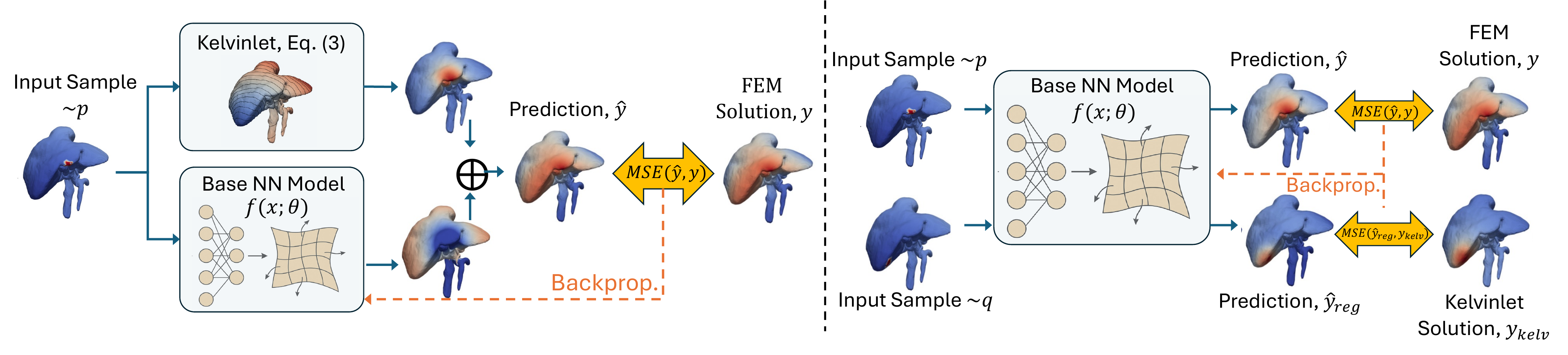}
    \caption{Overview of the two proposed training frameworks. \textbf{Method 1 (Left):} Residual learning predicts only the deviation from the Kelvinlet solution, reducing learning complexity by leveraging the physics-based analytical solution as an initialization. \textbf{Method 2 (Right):} Kelvinlet-based regularization combines FEM-supervised learning with a secondary loss enforcing consistency with Kelvinlet priors. The heatmap depicts the displacement norms.}


    \label{fig:methods}
\end{figure*}

The displacement field \( \mathbf{u}: \mathbb{R}^3 \to \mathbb{R}^3 \) of an isotropic, homogeneous elastic medium under an applied body force density \( \mathbf{f}: \mathbb{R}^3 \to \mathbb{R}^3 \) satisfies the Navier-Cauchy equations:
\begin{equation}
    \mu \nabla^2 \mathbf{u} + \frac{\mu}{1 - 2\nu} \nabla(\nabla \cdot \mathbf{u}) + \mathbf{f} = 0,
\end{equation}
where \( \mu > 0 \) is the shear modulus, and \( \nu \in (0, 0.5) \) is the Poisson's ratio, ensuring a physically consistent material response. Here, \( \nabla \) denotes the gradient operator, 
\( \nabla \cdot \mathbf{u} \) is the divergence of the displacement field, 
and \( \nabla^2 \mathbf{u} = \nabla \cdot (\nabla \mathbf{u}) \) is the vector Laplacian applied to \( \mathbf{u} \).

For a concentrated force \( \mathbf{f} \delta(\mathbf{x} - \mathbf{x}_s) \) applied at a source point \( \mathbf{x}_s \in \mathbb{R}^3 \), the Green’s function solution, known as the \emph{Kelvinlet}, provides an analytical displacement field \( \mathbf{u}: \mathbb{R}^3 \setminus \{\mathbf{x}_s\} \to \mathbb{R}^3 \) following:
\begin{equation} \label{eq:kelvinlet}
\begin{split}
    \mathbf{u}(\mathbf{x}) = \frac{1}{4 \pi \mu} \Bigg( 
    \frac{(1 - \nu) I}{r} 
    + \nu \frac{(\mathbf{x} - \mathbf{x}_s) \otimes (\mathbf{x} - \mathbf{x}_s)}{r^3} 
    \Bigg) \mathbf{f}
\end{split}
\end{equation}
where \( r = \| \mathbf{x} - \mathbf{x}_s \|_2 \) denotes the Euclidean distance from the source, and $\otimes$ denotes the outer product operator. 


The classical Kelvinlet in Eq.~\eqref{eq:kelvinlet} exhibits singularities at \( \mathbf{x}_s \). To prevent numerical instabilities, we substitute $r$ with $r_\varepsilon = \sqrt{\| \mathbf{x} - \mathbf{x}_s \|^2 + \varepsilon^2}$ for $\epsilon>0$,
where the spatial extent of force distribution is controlled by \( \varepsilon \), ensuring the solution is well-conditioned.
Finally, a re-scaled solution with respect to the local displacement boundary condition $\mathbf{u}_s=\mathbf{u}(\mathbf{x}_s)$ removes the dependence on both the force $\mathbf{f}$ and the modulus $\mu$, yielding the regularized Kelvinlet formulation used in this paper,
{\small
\begin{equation}
\mathbf{u}_\varepsilon(\mathbf{x}) 
= \frac{\varepsilon}{(5 - 6\nu)} \Bigg( 
    \frac{(3 - 4\nu) I}{r_\varepsilon} 
    + \frac{(\mathbf{x} - \mathbf{x}_s) \otimes (\mathbf{x} - \mathbf{x}_s)}{r_\varepsilon^3} 
\Bigg) \mathbf{u}_s.
\label{eq:singleKelvin}
\end{equation}}

\noindent\textbf{For multiple interaction points}, we note that at each interaction point we only observe the superposition of contributions from all interaction points, without knowing the exact contribution of each individual point to this superposition. Consequently, unlike the single-grasper setting, $\mathbf{u}_s$ in Eq.~\eqref{eq:singleKelvin} is not directly available for each interaction point and must be estimated. To formalize this, let $\mathbf{x}_i$ denote the location of the $i$th interaction, $\mathbf{u}_i$ the corresponding total displacement at $\mathbf{x}_i$, and $\mathbf{k}_i$ the unknown contribution of the $i$th interaction to this displacement. Assuming there are $K$ interaction points, the total displacement can be expressed as:

{\small
\begin{equation}
\mathbf{u}_\epsilon(\mathbf{x}) = \sum_{i=1}^K \underbrace{\frac{\varepsilon}{(5-6\nu)} \left( \frac{(3 - 4\nu) I}{r_\varepsilon} +  \frac{(\mathbf{x} - \mathbf{x}_i)\otimes(\mathbf{x} - \mathbf{x}_i)}{r_\varepsilon^3} \right)}_{\Gamma_i(\mathbf{x})}\mathbf{k_i},
\end{equation}}
where $\Gamma_i(\mathbf{x})\in \mathbb{R}^{3\times 3}$. and $r_{i,\varepsilon}(\mathbf{x}) := \sqrt{\|\mathbf{x}-\mathbf{x}_i\|_2^2 + \varepsilon^2}.$ Note that this formulation enables the total displacement function to incorporate cross communication between interaction points.

To estimate the $\mathbf{k}_i$ values, we solve the following optimization problem, which enforces the boundary conditions \( \mathbf{u}_\varepsilon(\mathbf{x}_i) = \mathbf{u}_i \) for all \(i\):  
\begin{equation}
    \operatorname*{argmin}_{\mathbf{k}} \sum_{i=1}^K \left( \| \sum_{j=1}^K \Gamma_j(\mathbf{x}_i)\mathbf{k}_j - \mathbf{u}_i \|^2) + \lambda \|\mathbf{k}_i\|^2\right).
    \label{eq:multiKelvin}
\end{equation}
This formulation ensures that the individual Kelvinlet contributions collectively satisfy the prescribed displacements, while the regularization term \(\lambda = 0.001\) promotes a minimum-energy solution. The optimization in Eq. \eqref{eq:multiKelvin} is essentially a regularized least squares problem, which admits a fast analytic solution:
\begin{equation}
    \mathbf{k}^* = 
    \left( \mathbf{\Gamma}^\top \mathbf{\Gamma} + \lambda \mathbf{I} \right)^{-1} 
    \mathbf{\Gamma}^\top \mathbf{u},
    \label{eq:multiKelvinClosedForm}
\end{equation}
where \(\mathbf{I}\in \mathbb{R}^{3K\times 3K}\) is the identity matrix, and $\mathbf{\Gamma}\in \mathbb{R}^{3K\times 3K}$, $\mathbf{k}\in \mathbb{R}^{3K}$, and $\mathbf{u}\in\mathbb{R}^{3K}$ are defined as:
\[
    \mathbf{\Gamma} = 
    \begin{bmatrix}
        \Gamma_1(\mathbf{x}_1) & \cdots & \Gamma_K(\mathbf{x}_1) \\
        \vdots & \ddots & \vdots \\
        \Gamma_1(\mathbf{x}_K) & \cdots & \Gamma_K(\mathbf{x}_K)
    \end{bmatrix},
    \mathbf{k} = 
    \begin{bmatrix}
        \mathbf{k}_1 \\ \vdots \\ \mathbf{k}_K
    \end{bmatrix},
    \mathbf{u} = 
    \begin{bmatrix}
        \mathbf{u}_1 \\ \vdots \\ \mathbf{u}_K
    \end{bmatrix}.
\]


Note that while the closed form solution requires the inversion of a $3K\times 3K$ matrix, the number of interaction points, $K$ is often small, leading to a very fast solution for Kelvinlet. The total Kelvinlet deformation model, including the analytic solution to the optimization, can be performed in less than \emph{1 ms} for two simultaneously engaged graspers (i.e., $K\approx 20$) using native GPU acceleration in PyTorch on an NVIDIA A6000 GPU and on a mesh with $\sim 10\text{k}$ nodes.

Having established a fast physics-based solution, we now focus on integrating Kelvinlets into our models.

\subsection{Method 1: Kelvinlet-Based Residual Learning}

For an initial displacement  \(\mathbf{u}_s\) located at \(\mathbf{x}_s\), let \( \mathbf{u}_{\text{true}}(\cdot;\mathbf{x}_s,\mathbf{u}_s) \) denote the ground-truth displacement field. Instead of directly regressing \( \mathbf{u}_{\text{true}} \), we introduce a residual learning framework where a neural network estimates the deviation from the Kelvinlet prior, which serves as an initial approximation. Specifically, we define the residual displacement as  
\(
\mathbf{r}(\mathbf{x};\mathbf{x}_s,\mathbf{u}_s) = \mathbf{u}_{\text{true}}(\mathbf{x};\mathbf{x}_s,\mathbf{u}_s) - \mathbf{u}_\varepsilon(\mathbf{x};\mathbf{x}_s,\mathbf{u}_s),
\)
where \( \mathbf{u}_\varepsilon \) represents the Kelvinlet solution. The neural network, parameterized by \( \boldsymbol{\theta} \in \mathbb{R}^d \), learns to approximate this residual via  
\(
\hat{\mathbf{r}}(\mathbf{x};\mathbf{x}_s,\mathbf{u}_s,\boldsymbol{\theta}).
\)
The corrected displacement field is then given by  
\(
\tilde{\mathbf{u}}(\mathbf{x};\mathbf{x}_s,\mathbf{u}_s,\boldsymbol{\theta}) = \mathbf{u}_\varepsilon(\mathbf{x};\mathbf{x}_s,\mathbf{u}_s) + \hat{\mathbf{r}}(\mathbf{x};\mathbf{x}_s,\mathbf{u}_s,\boldsymbol{\theta}).
\)
Let \(
\| \mathbf{u} \|_{L_2(\Omega)}^2 = \int_\Omega \|\mathbf{u}(\mathbf{x})\|^2 d\mathbf{x}
\), then to train the network, we minimize the loss:
{
\begin{multline}
\mathcal{L}_{\text{residual}}(\boldsymbol{\theta}) 
= \mathbb{E}_{(\mathbf{x}_s, \mathbf{u}_s) \sim p} \Big[ 
\| \mathbf{u}_{\text{true}}(\cdot;\mathbf{x}_s, \mathbf{u}_s) \\
- \tilde{\mathbf{u}}(\cdot;\mathbf{x}_s, \mathbf{u}_s,\boldsymbol{\theta}) 
\|^2_{L_2(\Omega)} \Big],
\end{multline}
}

where \( p \) represents the distribution of all possible pointwise interactions with the tissue, and \( \Omega \) denotes the tissue domain. By focusing on learning only the correction beyond the Kelvinlet solution, this approach leverages the physics priors directly into its formulation.

\subsection{Method 2: Kelvinlets as Regularization}

Alternatively, a network can be trained to directly regress \( \mathbf{u}_{\text{true}} \), producing an estimate \( \hat{\mathbf{u}} \). To enforce physically consistent predicted deformations by the network, we leverage Kelvinlets as a regularization prior and write: 
{
\begin{multline} \label{eq:full-loss}
\mathcal{L}(\boldsymbol{\theta}) = 
\mathbb{E}_{(\mathbf{x}_s, \mathbf{u}_s) \sim p} \left[ 
\left\| \mathbf{u}_{\text{true}} - \hat{\mathbf{u}} \right\|^2_{L_2(\Omega)} 
\right] \\
+ \lambda_{\text{reg}} \, \mathbb{E}_{(\mathbf{x}'_s, \mathbf{u}'_s) \sim q} \left[ 
\left\| \mathbf{u}_\varepsilon - \hat{\mathbf{u}} \right\|^2_{L_2(\Omega)} 
\right]
\end{multline}
}

where \( \lambda_{\text{reg}} > 0 \) is the regularization coefficient. For clarity and brevity, we omit the dependencies \( (\cdot;\mathbf{x}_s,\mathbf{u}_s) \) and \( (\cdot;\mathbf{x}_s,\mathbf{u}_s,\boldsymbol{\theta}) \) in the notation. Note that we explicitly separate \( (\mathbf{x}_s, \mathbf{u}_s) \) from \( (\mathbf{x}_s', \mathbf{u}'_s) \) because, in practice, the two distributions $p$ and $q$ could differ, and also while Monte Carlo integration is used for both terms, the first term requires FEM simulations to obtain \( \mathbf{u}_{\text{true}} \) (which is computationally expensive), whereas the second term only involves Kelvinlet evaluations (much faster). We will describe the distributions $p$ and $q$ used in our experiments in the following section. 

Figure ~\ref{fig:methods} illustrates the proposed residual- and regularization-based learning methods, along with the displacement norm visualizations for all methods on a representative sample.

\section{Experiments}

\subsubsection{Data.}
\label{sec:data}

\begin{figure*}[t]
    \centering
    \includegraphics[width=.9\textwidth]{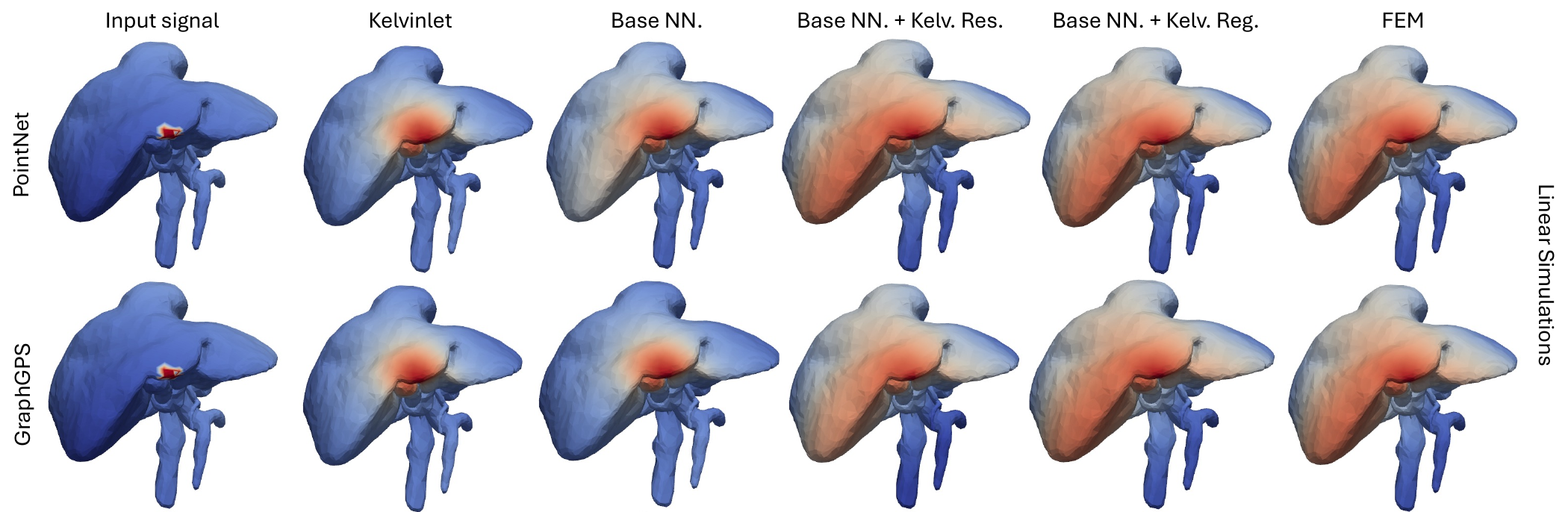}
    \caption{Qualitative comparison of deformation predictions for the best (PointNet) and worst (GraphGPS) models in linear regimes.}
    \label{fig:figure1}
\end{figure*}

Our dataset consists of large-scale FEM simulations of soft tissue deformation. Each node in the mesh has a feature vector: $\mathbf{f} = \left[\mathbf{x}; \mathbf{u}_s; a\right] \in \mathbb{R}^7$. Where $\mathbf{x} \in \mathbb{R}^3$ is the undeformed position, $\mathbf{u}_s \in \mathbb{R}^3$ is the initial displacement, and $a \in \{0,1\}$ indicates whether the node is actively influenced by external forces. This distinction helps differentiate constrained nodes from passively deforming ones. We create a dataset of 20,800 FEM solutions with 10,400 linear elastic simulations and 10,400 nonlinear material responses. Of these, 5,000 correspond to single manipulations and 5,400 to multiple grasper manipulations.

\subsubsection{Neural Architectures.}
We evaluate two groups of models: deformation-specific FEM surrogates and generic set/graph architectures. For the former, PhysGNN \cite{salehi2022physgnn} is a mesh-based GNN that we augment with a global attention layer to counter oversmoothing \cite{zhao2020pairnorm}; PEGNN \cite{saleh2024physics} is used only through its soft-tissue encoder; SeeSaw \cite{10598377} is reimplemented for our point-cloud setting; from the diffeomorphic framework of \cite{10630572} we keep only the deformation network; and a Transformer-based KAN \cite{JIANG2025126619} is restricted to single-step prediction. To study architecture-agnostic effects of Kelvinlet priors, we also use U-Net \cite{mendizabal2020simulation}, SAB \cite{lee2019set}, GraphGPS \cite{rampasek2022GPS}, and PointNet \cite{qi2016pointnet}: U-Net is converted into a DeepSet-style model via shared pointwise MLPs, SAB uses two self-attention blocks with an MLP head, and GraphGPS and PointNet are shallow variants with reduced depth to avoid overfitting on high-resolution meshes.

\begin{figure}[t!]
\vspace{-0.15in}
    \centering
    \includegraphics[width=0.45\textwidth]{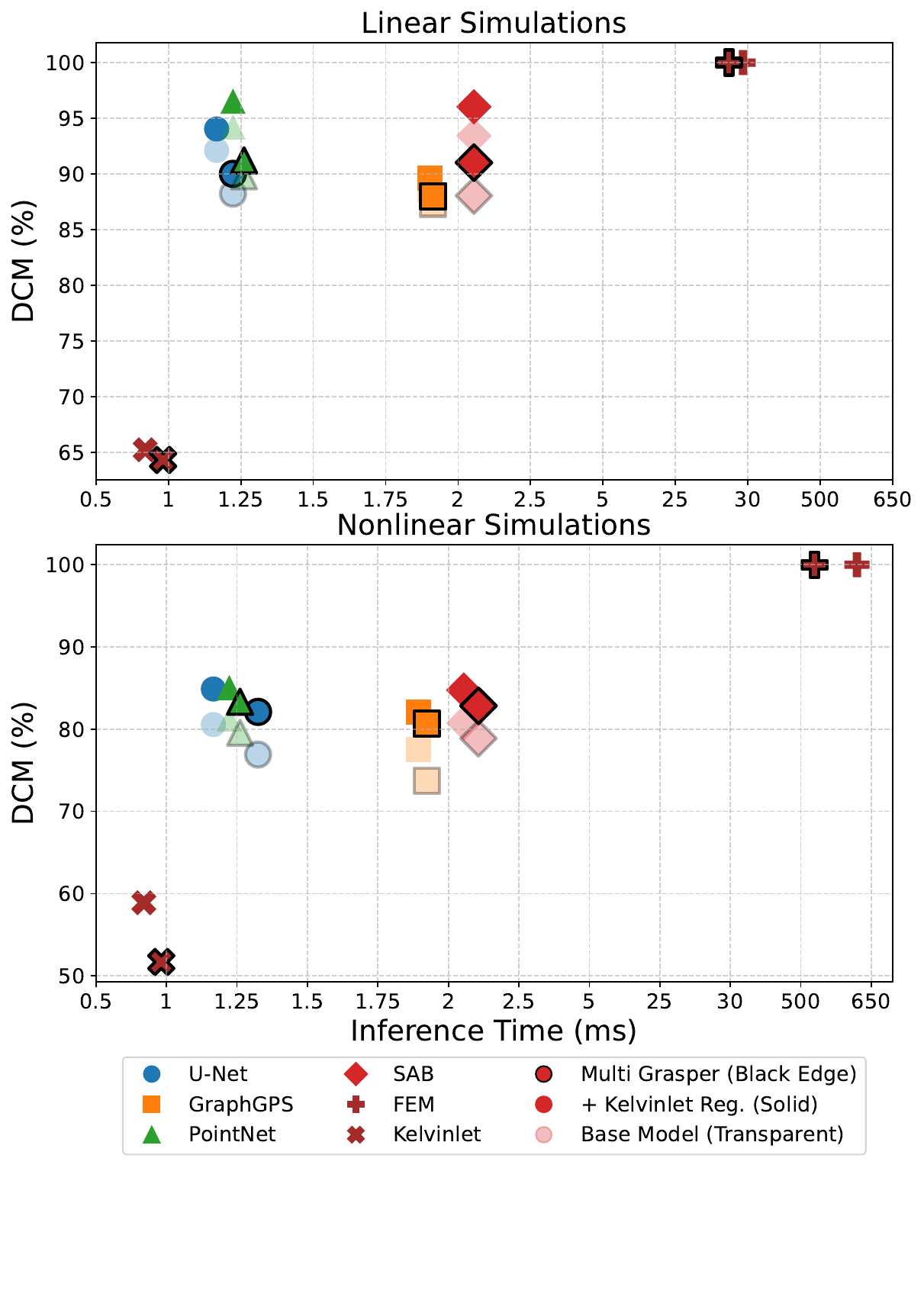}
    \vspace{-0.45in}
    \caption{Log-scaled inference time vs. accuracy (DCM\%) for various models in linear and nonlinear simulations. Transparent markers denote base models, while solid markers indicate Kelvinlet-regularized versions. Black-edged markers represent multi-grasper cases, while edge-free markers denote single-grasper cases. FEM and Kelvinlet baselines are included for reference.}
    \label{fig:runtimes}
    \vspace{-0.2in}
\end{figure}

\subsubsection{Hyperparameters.}
\label{subsec:hyperparameters}

We set shear modulus in Kelvinlet to $\mu = 0.72$ kPa, Poisson’s ratio $\nu = 0.45$, and regularization parameter $\epsilon = 0.05$, to reflect soft tissue properties and ensure numerical stability. For linear and nonlinear simulations we use $\lambda_{\text{reg}} = 1$ and $\lambda_{\text{reg}} = 0.1$, respectively, to emphasize linearity of Kelvinlet.

\subsubsection{Results and Analysis}
\label{subsec:results}

\begin{table}[t]
    \scriptsize
    \centering
    \renewcommand{\arraystretch}{1.1}
    \setlength{\tabcolsep}{3pt}

    \scalebox{0.91}{%
    \begin{tabular}{lccc}
        \toprule
        \multirow{2}{*}{\textbf{Model}} 
        & \textbf{Base} 
        & \multicolumn{2}{c}{\textbf{Ours}} \\
        \cmidrule(lr){3-4}
        &  & \textbf{+Res.} & \textbf{+Reg.} \\
        \midrule
        PhysGNN\cite{salehi2022physgnn}              & 1.71 (0.11) & 1.84 (0.13) & 1.71 (0.11) \\
        SeeSaw\cite{10598377}                       & 1.56 (0.09) & 1.69 (0.08) & 1.56 (0.09) \\
        Physics-Encoded GNN\cite{saleh2024physics}   & 1.65 (0.08) & 1.79 (0.08) & 1.65 (0.08) \\
        Transformer-KAN\cite{JIANG2025126619}        & 2.22 (0.17) & 2.32 (0.16) & 2.22 (0.17) \\
        Diffeomorphic Deformation\cite{10630572}     & 2.16 (0.12) & 2.35 (0.11) & 2.16 (0.12) \\
        PointNet\cite{qi2016pointnet}               & 1.16 (0.10) & 1.23 (0.13) & 1.16 (0.10) \\
        U-Net\cite{mendizabal2020simulation}        & 1.55 (0.08) & 1.64 (0.07) & 1.55 (0.08) \\
        GraphGPS\cite{rampasek2022GPS}              & 2.60 (0.15) & 2.78 (0.15) & 2.60 (0.15) \\
        SAB\cite{lee2019set}                        & 2.30 (0.02) & 2.44 (0.21) & 2.30 (0.20) \\
        \midrule
        Kelvinlet                                   & \textbf{0.15 (0.052)} & \multicolumn{2}{c}{ } \\
        \bottomrule
    \end{tabular}%
    }
    \caption{Inference time (ms) on an NVIDIA RTX A6000 GPU for a single grasper in linear simulations, with and without Kelvinlet priors. Values are mean (std).}
    \label{tab:inferenceTimeTable}
    \vspace{-0.1in}
\end{table}

\begin{table*}[t]
    \centering
    \renewcommand{\arraystretch}{1.5}
    \setlength{\tabcolsep}{5pt}
    \scalebox{0.68}{%
    \begin{tabular}{l|ccc|ccc|ccc}
        \toprule
        \multirow{2}{*}{\textbf{Model}} 
        & \multicolumn{3}{c|}{\textbf{Indiv. (DCM)}} 
        & \multicolumn{3}{c|}{\textbf{Comb. (DCM)}} 
        & \multicolumn{3}{c}{\textbf{Multi (DCM)}} \\
        \cmidrule(lr){2-4} \cmidrule(lr){5-7} \cmidrule(lr){8-10}
        & Base & +Res. & +Reg. 
        & Base & +Res. & +Reg. 
        & Base & +Res. & +Reg. \\
        \midrule
        PhysGNN\cite{salehi2022physgnn}              
            & 90.12(0.13) & 91.77(0.11) & 92.65(0.10)
            & 86.38(0.13) & 88.10(0.11) & 88.66(0.10)
            & 89.55(0.13) & 90.96(0.11) & 91.54(0.10) \\
        SeeSaw\cite{10598377}                        
            & 89.24(0.15) & 91.02(0.10) & 91.83(0.08)
            & 85.50(0.15) & 87.35(0.10) & 87.84(0.08)
            & 88.67(0.15) & 90.21(0.10) & 90.72(0.08) \\
        Physics-Encoded GNN\cite{saleh2024physics}   
            & 88.55(0.14) & 90.44(0.09) & 91.58(0.12)
            & 84.81(0.14) & 86.77(0.09) & 87.59(0.12)
            & 87.98(0.14) & 89.63(0.09) & 90.47(0.12) \\
        Transformer-KAN\cite{JIANG2025126619}        
            & 90.89(0.12) & 92.01(0.08) & 93.22(0.07)
            & 87.15(0.12) & 88.34(0.08) & 89.23(0.07)
            & 90.32(0.12) & 91.20(0.08) & 92.11(0.07) \\
        Diffeomorphic Deformation\cite{10630572}      
            & 91.14(0.10) & 92.88(0.11) & 94.33(0.09)
            & 87.40(0.10) & 89.21(0.11) & 90.34(0.09)
            & 90.57(0.10) & 92.07(0.11) & 93.22(0.09) \\
        PointNet\cite{qi2016pointnet}     
            & 94.23(0.11) & 94.66(0.09) & \textbf{96.54(0.09)} 
            & 89.79(0.07) & 90.11(0.07) & \textbf{91.22(0.07)} 
            & 94.00(0.08) & 94.12(0.07) & \textbf{95.01(0.06)} \\
        U-Net\cite{mendizabal2020simulation}    
            & 92.11(0.12) & 93.56(0.10) & 94.04(0.11) 
            & 88.23(0.08) & 89.67(0.09) & 90.01(0.08) 
            & 91.20(0.09) & 92.33(0.10) & 92.88(0.08) \\
        GraphGPS\cite{rampasek2022GPS}    
            & 88.60(0.21) & 89.45(0.15) & 89.61(0.23) 
            & 87.33(0.06) & 87.92(0.02) & 88.01(0.04) 
            & 88.35(0.13) & 88.91(0.10) & 89.45(0.09) \\
        SAB\cite{lee2019set}       
            & 93.42(0.11) & 94.60(0.04) & 96.03(0.04) 
            & 88.05(0.09) & 89.91(0.11) & 91.02(0.04) 
            & 92.54(0.06) & 93.66(0.07) & 94.42(0.06) \\
        \midrule
        Kelvinlet 
            & \multicolumn{3}{c|}{65.24(0)} 
            & \multicolumn{3}{c|}{64.33(0)} 
            & \multicolumn{3}{c}{64.82(0)} \\
        \bottomrule
    \end{tabular}%
    }
    \caption{Performance of various models on \textbf{linear} simulations with individual graspers, combined graspers, and multitask learning. DCM (\%) is reported with standard deviations. ``Base'' denotes the original model, ``+Res.'' augments it with a Kelvinlet-based residual, and ``+Reg.'' applies a Kelvinlet-inspired regularizer.}
    \label{tab:resultsLinearMerged}
\end{table*}

\begin{table*}[t]
    \centering
    \renewcommand{\arraystretch}{1.5}
    \setlength{\tabcolsep}{5pt}
    \scalebox{0.68}{%
    \begin{tabular}{l|ccc|ccc|ccc}
        \toprule
        \multirow{2}{*}{\textbf{Model}} 
        & \multicolumn{3}{c|}{\textbf{Indiv. (DCM)}} 
        & \multicolumn{3}{c|}{\textbf{Comb. (DCM)}} 
        & \multicolumn{3}{c}{\textbf{Multi (DCM)}} \\
        \cmidrule(lr){2-4} \cmidrule(lr){5-7} \cmidrule(lr){8-10}
        & Base & +Res. & +Reg. 
        & Base & +Res. & +Reg. 
        & Base & +Res. & +Reg. \\
        \midrule
        PhysGNN\cite{salehi2022physgnn}              
            & 78.05(0.13) & 82.19(0.11) & 82.79(0.10)
            & 75.30(0.13) & 79.18(0.11) & 80.83(0.10)
            & 77.73(0.13) & 80.58(0.11) & 81.34(0.10) \\
        SeeSaw\cite{10598377}                        
            & 77.17(0.15) & 81.44(0.10) & 81.97(0.08)
            & 74.42(0.15) & 78.43(0.10) & 80.01(0.08)
            & 76.85(0.15) & 79.83(0.10) & 80.52(0.08) \\
        Physics-Encoded GNN\cite{saleh2024physics}   
            & 76.48(0.14) & 80.86(0.09) & 81.72(0.12)
            & 73.73(0.14) & 77.85(0.09) & 79.76(0.12)
            & 76.16(0.14) & 79.25(0.09) & 80.27(0.12) \\
        Transformer-KAN\cite{JIANG2025126619}        
            & 78.82(0.12) & 82.43(0.08) & 83.36(0.07)
            & 76.07(0.12) & 79.42(0.08) & 81.40(0.07)
            & 78.50(0.12) & 80.82(0.08) & 81.91(0.07) \\
        Diffeomorphic Deformation\cite{10630572}      
            & 79.07(0.10) & 83.30(0.11) & 84.47(0.09)
            & 76.32(0.10) & 80.29(0.11) & 82.51(0.09)
            & 78.75(0.10) & 81.69(0.11) & 83.02(0.09) \\
        PointNet\cite{qi2016pointnet}     
            & 81.25(0.22) & 84.65(0.19) & \textbf{85.05(0.15)} 
            & 79.56(0.18) & 82.66(0.16) & \textbf{83.33(0.11)} 
            & 83.25(0.17) & 84.75(0.13) & \textbf{85.21(0.09)} \\
        U-Net\cite{mendizabal2020simulation}    
            & 80.56(0.10) & 84.22(0.09) & 84.88(0.11) 
            & 76.91(0.05) & 79.83(0.08) & 82.10(0.21) 
            & 78.10(0.12) & 80.45(0.09) & 81.33(0.14) \\
        GraphGPS\cite{rampasek2022GPS}    
            & 77.52(0.17) & 80.63(0.13) & 82.11(0.09) 
            & 73.72(0.15) & 77.92(0.10) & 80.66(0.16) 
            & 76.24(0.18) & 79.31(0.15) & 80.42(0.10) \\
        SAB\cite{lee2019set}       
            & 80.75(0.19) & 84.44(0.07) & 84.75(0.10) 
            & 78.88(0.11) & 81.50(0.15) & 82.83(0.16) 
            & 81.22(0.10) & 83.01(0.08) & 84.00(0.07) \\
        \midrule
        Kelvinlet 
            & \multicolumn{3}{c|}{58.88(0)} 
            & \multicolumn{3}{c|}{51.67(0)} 
            & \multicolumn{3}{c}{55.56(0)} \\
        \bottomrule
    \end{tabular}%
`    }
    \caption{Performance of various models on \textbf{nonlinear} simulations with individual graspers, combined graspers, and multitask learning. DCM (\%) is reported with standard deviations. ``Base'' denotes the original model, ``+Res.'' augments it with a Kelvinlet-based residual, and ``+Reg.'' applies a Kelvinlet-inspired regularizer.}
    \label{tab:resultsNonLinearMerged}
\end{table*}

\noindent To quantitatively assess model performance, we utilize the \textit{Deformation Capture Mean} (DCM) metric, which measures how well a predicted deformation field approximates the ground-truth displacement field. The DCM is computed as:
{\small
\begin{equation}
    \text{DCM} = 100 \times 
    \left( 
    1 - \mathbb{E}_{(\mathbf{x}_s, \mathbf{u}_s)\sim p}
    \left[ 
    \frac{\left\|\mathbf{u}_{\text{true}} - \hat{\mathbf{u}}\right\|_{L_2(\Omega)}}{\|\mathbf{u}_{\text{true}}\|_{L_2(\Omega)}} 
    \right] 
    \right)
\end{equation}}

\noindent Table \ref{tab:resultsLinearMerged} shows that Kelvinlet priors consistently improve deformation accuracy in the linear regime. Across all architectures, residual learning yields roughly 1 to 2 percentage points of DCM improvement, while Kelvinlet regularization adds a further 1 to 2 points, with the largest gain of \textbf{+3.19\%} achieved by the diffeomorphic deformation network. PointNet achieves the highest absolute performance, reaching \textbf{96.54\%} DCM in the single grasper setting. In contrast, the analytical Kelvinlet alone remains substantially less accurate at \textbf{65.24\%} DCM, highlighting the importance of learned residual corrections.

As shown in Table~\ref{tab:inferenceTimeTable}, the proposed methods incur negligible runtime overhead. Regularized models match the base inference time, while residual variants increase latency by only about 0.1ms. Although standalone Kelvinlet inference is extremely fast at \textbf{0.15ms}, its accuracy is insufficient, supporting its role as a prior rather than a predictor. Overall, accuracy gains are achieved without sacrificing efficiency.

Figure~\ref{fig:runtimes} illustrates the accuracy versus runtime trade off across linear and nonlinear simulations. Kelvinlet regularization consistently improves accuracy relative to the base models, as indicated by the upward shift of solid markers, while maintaining comparable inference times. Models trained on multi grasper data follow the same trend, indicating robust generalization across grasping conditions. All runtime measurements were obtained on a single NVIDIA A6000 GPU.
\vspace{-0.15in}
\section{Conclusion}
\label{sec:conclusion}

We present Neural Augmented Kelvinlets, a physics informed framework for real time soft tissue deformation modeling that improves accuracy, physical plausibility, and generalization while preserving real time performance. Kelvinlets are incorporated both as a residual correction mechanism and as a regularization prior that enforces physically consistent deformations. Extensive FEM simulations demonstrate clear performance gains, particularly in multi grasper scenarios. Our GPU accelerated PyTorch implementation enables real time inference and realistic visualization, making the method suitable for surgical robotics and medical training applications.

\section*{Acknowledgments}
This work was fully supported by the Wellcome-Leap SAVE program.

\bibliography{aaai2026}

\newpage
\clearpage

\setlength{\leftmargini}{20pt}
\makeatletter\def\@listi{\leftmargin\leftmargini \topsep .5em \parsep .5em \itemsep .5em}
\def\@listii{\leftmargin\leftmarginii \labelwidth\leftmarginii \advance\labelwidth-\labelsep \topsep .4em \parsep .4em \itemsep .4em}
\def\@listiii{\leftmargin\leftmarginiii \labelwidth\leftmarginiii \advance\labelwidth-\labelsep \topsep .4em \parsep .4em \itemsep .4em}\makeatother

\setcounter{secnumdepth}{0}
\renewcommand\thesubsection{\arabic{subsection}}
\renewcommand\labelenumi{\thesubsection.\arabic{enumi}}

\newcounter{checksubsection}
\newcounter{checkitem}[checksubsection]

\newcommand{\checksubsection}[1]{%
  \refstepcounter{checksubsection}%
  \paragraph{\arabic{checksubsection}. #1}%
  \setcounter{checkitem}{0}%
}

\newcommand{\checkitem}{%
  \refstepcounter{checkitem}%
  \item[\arabic{checksubsection}.\arabic{checkitem}.]%
}
\newcommand{\question}[2]{\normalcolor\checkitem #1 #2 \color{blue}}
\newcommand{\ifyespoints}[1]{\makebox[0pt][l]{\hspace{-15pt}\normalcolor #1}}

\end{document}